\documentclass[10pt,twocolumn,twoside]{IEEEtran}
\usepackage[utf8]{inputenc}
\usepackage{amsmath,authblk,acronym,cite}
\usepackage{graphicx}
\usepackage{subcaption}

\begin{document}

\newacronym{DTW}{Dynamic Time Warping}
\newacronym{FFT}{Fast Fourier Transform}
\newacronym{GRASSHOPPER}{Graph Random-walk with Absorbing StateS that HOPs
among PEaks for Ranking}
\newacronym{HMM}{Hidden Markov Model}
\newacronym{ISMIR}{International Society for Music Information Retrieval}
\newacronym{KL}{Kullback-Leibler}
\newacronym{LSA}{Latent Semantic Analysis}
\newacronym{MFCC}{Mel Frequency Cepstral Coefficient}
\newacronym{MIREX}{Music Information Retrieval Evaluation eXchange}
\newacronym{MIR}{Music Information Retrieval}
\newacronym{MMR}{Maximal Marginal Relevance}
\newacronym{pp}{percentage points}
\newacronym{RMS}{Root Mean Square}
\newacronym{SVD}{Singular Value Decomposition}
\newacronym{SVM}{Support Vector Machine}

\title{Using Generic Summarization to Improve Music Information Retrieval Tasks}
\author{Francisco Raposo, Ricardo Ribeiro, David Martins de
Matos,~\IEEEmembership{Member,~IEEE} \thanks{F. Raposo and D. Martins de Matos
are with Instituto Superior Técnico, Universidade de Lisboa, Av. Rovisco Pais,
1049-001 Lisboa, Portugal. R. Ribeiro is with Instituto Universitário de Lisboa
(ISCTE-IUL), Av. das Forças Armadas, 1649-026 Lisboa, Portugal. F. Raposo, R.
Ribeiro, and D. Martins de Matos are with INESC-ID Lisboa, R. Alves Redol 9,
1000-029 Lisboa, Portugal. This work was supported by national funds
through Fundação para a Ciência e a Tecnologia (FCT) with reference
UID/CEC/50021/2013.}}
\maketitle

\begin{abstract}
In order to satisfy processing time constraints, many \MIR~tasks process only a
segment of the whole music signal. This may lead to decreasing performance, as
the most important information for the tasks may not be in the processed
segments. We leverage generic summarization algorithms, previously applied to
text and speech, to summarize items in music datasets. These algorithms build
summaries (both concise and diverse), by selecting appropriate segments from the
input signal, also making them good candidates to summarize music. We evaluate
the summarization process on binary and multiclass music genre classification
tasks, by comparing the accuracy when using summarized datasets against the
accuracy when using human-oriented summaries, continuous segments (the
traditional method used for addressing the previously mentioned time
constraints), and full songs of the original dataset. We show that
\GRASSHOPPERshort, LexRank, \LSAshort, \MMRshort, and a Support Sets-based
centrality model improve classification performance when compared to selected
baselines. We also show that summarized datasets lead to a classification
performance whose difference is not statistically significant from using full
songs. Furthermore, we make an argument stating the advantages of sharing
summarized datasets for future \MIR~research.
\end{abstract}

\section{Introduction}
Music summarization has been the subject of research for at least a decade and
many algorithms that address this problem, mainly for popular music, have been
published in the past
\cite{Chai2006,Cooper2003,Peeters2002,Peeters2003,Chu2000,Cooper2002,Glaczynski2011,Bartsch2005}.
However, those algorithms focus on producing human consumption-oriented
summaries, i.e., summaries that will be listened to by people motivated by the
need to quickly get the gist of the whole song without having to listen to all
of it. This type of summarization entails extra requirements besides conciseness
and diversity (non-redundancy), such as clarity and coherence, so that people
can enjoy listening to them.

Generic summarization algorithms, however, focus on extracting concise and
diverse summaries and have been successfully applied in text and speech
summarization \cite{Carbonell1998,Erkan2004,Landauer1997,Zhu2007,Ribeiro2011}.
Their application, in music, for human consumption-oriented purposes is not
ideal, for they will select and concatenate the most relevant and diverse
information (according to each algorithm's definition of relevance and
diversity) without taking into account whether the output is enjoyable for
people or not. This is usually reflected, for instance, on discontinuities or
irregularities in beat synchronization in the resulting summaries.

We focus on improving the performance of tasks recognized as important by the
\MIR~community, e.g. music genre classification, through summarization, as
opposed to considering music summaries as the product to be consumed by people.
Thus, we can ignore some of the requirements of previous music summarization
efforts, which usually try to model the musical structure of the pieces being
summarized, possibly using musical knowledge. Although human-related aspects of
music summarization are important in general, they are beyond the focus of this
paper. We claim that, for \MIR~tasks benefiting from summaries, it is sufficient
to consider the most relevant parts of the signal, according to its features. In
particular, summarizers do not need to take into account song structure or human
perception of music. Our rationale is that summaries contain more relevant and
less redundant information, thus improving the performance of tasks that rely on
processing just a portion of the whole signal, leading to faster processing,
less space usage, and efficient use of bandwidth.

We use \GRASSHOPPERshort~\cite{Zhu2007}, LexRank~\cite{Erkan2004},
\LSAshort~\cite{Landauer1997}, \MMRshort~\cite{Carbonell1998}, and Support
Sets~\cite{Ribeiro2011} to summarize music for automatic (instead of human)
consumption. To evaluate the effects of summarization, we assess the performance
of binary and 5-class music genre classification, when considering song
summaries against continuous clips (taken from the beginning, middle, and end of
the songs) and against the whole songs. We show that all of these algorithms
improve classification performance and are statistically not significantly
different from using the whole songs. These results complement and solidify
previous work evaluated on a binary Fado classifier~\cite{Raposo2015}.

The article is organized as follows: section \ref{sec:music-summarization}
reviews related work on music-specific summarization. Section
\ref{sec:generic-summarization} reviews the generic summarization algorithms we
experimented with: \GRASSHOPPERshort~(section \ref{sub:grasshopper}), LexRank
(section \ref{sub:lexrank}), \LSAshort~(section \ref{sub:lsa}),
\MMRshort~(section \ref{sub:mmr}), and Support Sets-based Centrality (section
\ref{sub:support-sets}). Section \ref{sec:experiments} details the experiments
we performed for each algorithm and introduces the classifier. Sections
\ref{sec:binary-results} and \ref{sec:multiclass-results} report our
classification results for the binary and multiclass classification scenarios,
respectively. Section \ref{sec:discussion} discusses the results and Section
\ref{sec:conclusions} concludes this paper with some remarks and future work.

\section{Music Summarization}\label{sec:music-summarization}
Current algorithms for music summarization were developed to extract an
enjoyable summary so that people can listen to it clearly and coherently.
In contrast, our approach considers summaries exclusively for automatic
consumption.

Human-oriented music summarization starts by structurally segmenting songs and
selecting meaningful segments to include in the summary. The assumption is that
songs are represented as label sequences where each label represents a different
part of the song (e.g., ABABCA where A is the chorus, B the verse, and C the
bridge). In \cite{Chai2006}, segmentation is achieved by using a \HMM~to detect
key changes between frames and \DTW~to detect repeating structure. In
\cite{Cooper2003}, a Gaussian-tempered ``checkerboard'' kernel is correlated
along the main diagonal of the song's self-similarity matrix, outputting segment
boundaries. Then, a segment-indexed matrix, containing the similarity between
detected segments, is built. \SVD~is applied to find its rank-$K$ approximation.
Segments are, then, clustered to output the song's structure. In
\cite{Peeters2002,Peeters2003}, a similarity matrix is
built and analyzed for fast changes, outputting segment boundaries; segments are clustered
to output the ``middle states''; an \HMM~is applied to these states,
producing the final segmentation. Then, various strategies are considered to
select the appropriate segments.

In \cite{Chu2000}, a modification of the \KL~divergence is used to group and
label similar segments. The summary consists of the longest sequence of segments
belonging to the same cluster. In \cite{Cooper2002} and \cite{Glaczynski2011},
Average Similarity is used to extract a thumbnail $L$ seconds long that is the
most similar to the whole piece. It starts by calculating a similarity matrix
through computing frame-wise similarities. Then, it calculates an aggregated
similarity measure, for each possible starting frame, of the $L$-second segment
with the whole song and picks the one that maximizes it as the summary. Another
method for this task, Maximum Filtered Correlation~\cite{Bartsch2005}, starts by
building a similarity matrix and then a filtered time-lag matrix, embedding the
similarity between extended segments separated by a constant lag. The starting
frame of the summary corresponds to the index that maximizes the filtered
time-lag matrix. In \cite{Xu2005}, music is classified as pure or vocal, in
order to perform type-specific feature extraction. The summary, created from
three to five seconds subsummaries (built from frame clusters), takes into
account musicological and psychological aspects, by differentiating between
types of music based on feature selection and specific duration. This promotes
human enjoyment when listening to the summary. Since these summaries were
targeted to people, they were evaluated by people.

In \cite{Vaizman2014}, music datasets are summarized into a codebook-based
audio feature representation, to efficiently retrieve songs in a
query-by-tag and query-by-example fashion. An initial dataset is
discretized, creating a dictionary of $k$ basis vectors. Then, for each query
song, the audio signal is quantized, according to the pre-computed
dictionary, mapping the audio signal into a histogram of basis
vectors. These histograms are used to compute music
similarity. This type of summarization allows for efficient retrieval of music
but is limited to the features which are initially chosen. Our focus is on audio
signal summaries, which are suitable for any audio feature extraction, instead
of proxy representations for audio features.

\section{Generic Summarization}\label{sec:generic-summarization}
Applying generic summarization to music implies song segmentation into musical
words and sentences. Since we do not take into account human-related aspects of
music perception, we can segment songs according to an arbitrarily fixed size.
This differs from structural segmentation in that it does not take into account
human perception of musical structure and does not create meaningful segments.
Nevertheless, it still allows us to look at the variability and repetition of
the signal and use them to find its most important parts. Furthermore, since it
is not aimed at human consumption, the generated summaries are less liable to
violate the copyrights of the original songs. This facilitates the sharing of
datasets (using the signal itself, instead of specific features extracted from
it) for \MIR~research efforts. In the following sections, we review the generic
summarization algorithms we evaluated.

\subsection{\GRASSHOPPERshort}\label{sub:grasshopper}
The \GRASSHOPPER~\cite{Zhu2007} was applied to text summarization and social
network analysis, focusing on improving ranking diversity. It takes an
$n{\times}n$ matrix $W$ representing a graph where each sentence is a vertex and
each edge has weight $w_{ij}$ corresponding to the similarity between sentences
$i$ and $j$; and a probability distribution $\boldsymbol{r}$ encoding prior
ranking.
First, $W$ is row-normalized: $O_{ij}{=}w_{ij}/\sum_{k=1}^{n}w_{ik}$.
Then, $P{=}\lambda O{+}(1{-}\lambda)\boldsymbol{1r}^{T}$ is built, incorporating
the user-supplied prior ranking $\boldsymbol{r}$ ($\boldsymbol{1}$ is an all-1
vector, $\boldsymbol{1r}^{T}$ is the outer product, and $\lambda$ is a balancing
factor).
The first ranking state $g_{1}{=}\operatorname{argmax}_{i=1}^{n}{\pi_{i}}$ is
found by taking the state with the largest stationary probability
($\pi{=}P^{T}\pi$ is the stationary distribution of $P$). Each time a state is
extracted, it is converted into an absorbing state to penalize states similar to
it. The rest of the states are iteratively selected according to the expected
number of visits to each state, instead of considering the stationary
probability. If $G$ is the set of items ranked so far, states are turned into
absorbing states by setting $P_{gg}{=}1$ and $P_{gi}{=}0$, $\forall{i{\neq} g}$.
If items are arranged so that ranked ones are listed before unranked ones, $P$
can be written as follows:
\begin{equation}
P=\begin{bmatrix}\boldsymbol{I}_{G} & \boldsymbol{0}\\
R & Q
\end{bmatrix}
\end{equation}
$\boldsymbol{I}_{G}$ is the identity matrix on $G$. $R$ and $Q$
are rows of unranked items. $N{=}(\boldsymbol{I}{-}Q)^{-1}$ is the expected
number of visits to state $j$ starting from state $i$ ($N_{ij}$). The expected
number of visits to state $j$, $v_{j}$, is given by
$\boldsymbol{v}{=}({N^{T}\boldsymbol{1}}){/}({n{-}|G|})$ and the next item is
$g_{|G|+1}{=}\operatorname{argmax}_{i=|G|+1}^{n}v_{i}$, where $|G|$ is the size of $G$.

\subsection{LexRank}\label{sub:lexrank}
LexRank \cite{Erkan2004} relies on the similarity (e.g. cosine) between sentence
pairs (usually, \emph{tf-idf} vectors). First, all sentences are compared to
each other. Then, a graph is built where each sentence is a vertex and edges are
created between every sentence according to their pairwise similarity (above a
threshold). LexRank can be used with both weighted (eq.~\ref{eq:lex-rank-w}) and
unweighted (eq.~\ref{eq:lex-rank-u}) edges. Then, each vertex score is
iteratively computed. In eq. \ref{eq:lex-rank-w} through \ref{eq:lex-rank-u},
$d$ is a damping factor to guarantee convergence; $N$ is the number of vertices;
$S\left(V_{i}\right)$ is the score of vertex $i$; and $D\left(V_{i}\right)$ is
the degree of $i$. Summaries are built by taking the highest ranked
sentences.

In LexRank, sentences recommend each other: sentences similar to many others
will get high scores. Scores are also determined by the score of the
recommending sentences.
\begin{equation}\label{eq:lex-rank-w}
S\left(V_{i}\right)=\frac{\left(1-d\right)}{N}+ S_{1}\left(V_i\right)
\end{equation}
\begin{equation}
S_{1}\left(V_{i}\right)=d\times\sum_{V_{j}\in
adj\left[V_{i}\right]}\frac{\text{Sim}\left(V_{i},V_{j}\right)}{\sum_{V_{k}\in
adj\left[V_{j}\right]}\text{Sim}\left(V_{j},V_{k}\right)}S\left(V_{j}\right)
\end{equation}
\begin{equation}\label{eq:lex-rank-u}
S\left(V_{i}\right)=\frac{\left(1-d\right)}{N}+d\times\sum_{V_{j}\in
adj\left[V_{i}\right]}\frac{S\left(V_{j}\right)}{D\left(V_{j}\right)}
\end{equation}

\subsection{\LSA}\label{sub:lsa}
\LSA~was first applied in text summarization in \cite{Gong2001}. \SVD~is used to
reduce the dimensionality of an original matrix representation of the text.
\LSA-based summarizers start by building a $T$ terms by $N$ sentences matrix
$A$. Each element of $A$, $a_{ij}{=}L_{ij}G_{i}$, has a local ($L_{ij}$) and a
global ($G_{i}$) weight. $L_{ij}$ is a function of term frequency in a specific
sentence and $G_{i}$ is a function of the number of sentences that contain a
specific term. Usually,  $a_{ij}$ are \emph{tf-idf} scores. The result of
applying the \SVD~to $A$ is $A{=}U\Sigma V^{T}$, where $U$ ($T{\times} N$
matrix) are the left singular vectors; $\Sigma$ ($N{\times} N$ diagonal matrix)
contains the singular values in descending order; and $V^{T}$ ($N{\times} N$
matrix) are the right singular vectors. Singular values determine topic
relevance: each latent dimension corresponds to a topic.
The rank-$K$ approximation considers the first $K$ columns of $U$, the
$K{\times} K$ sub-matrix of $\Sigma$, and the first $K$ rows of $V^{T}$.
Relevant sentences are the ones corresponding to the indices of the highest
values for each right singular vector. This approach has two limitations
\cite{Steinberger2004}: by selecting $K$ sentences for the summary, less
significant sentences tend to be extracted when $K$ increases; and, sentences
with high values in several topics, but never the highest, will
never be included in the summary. To account for these effects, a sentence
score was introduced and $K$ is chosen so that the $K^{th}$ singular value does
not fall under half of the highest singular value:
$score\left(j\right){=}\sqrt{\sum_{i=1}^{k}v_{ij}^{2}\sigma_{i}^{2}}$.

\subsection{\MMR}\label{sub:mmr}
Sentence selection in \MMR~\cite{Carbonell1998} is done according to their
relevance and diversity against previously selected sentences, in order to
output low-redundancy summaries. \MMR~is a query-based method that has been
used in speech summarization \cite{Zechner2000,Murray2005}. It is also possible
to produce generic summaries by taking the centroid vector of all the sentences
as the query.

\MMR~uses
$\lambda{\text{Sim}_{1}}\left(S_{i},Q\right){-}\left(1{-}\lambda\right)\max_{S_{j}}\text{Sim}_{2}\left(S_{i},S_{j}\right)$
to select sentences.
$\text{Sim}_{1}$~and~$\text{Sim}_{2}$ are similarity metrics (e.g. cosine);
$S_{i}$ and $S_{j}$ are unselected and previously selected sentences,
respectively; $Q$ is the query, and $\lambda$ balances relevance and diversity.
Sentences can be represented as \emph{tf-idf} vectors.

\subsection{Support Sets-based Centrality}\label{sub:support-sets}
This method was first applied in text and speech summarization
\cite{Ribeiro2011}. Centrality is based on sets of sentences that are similar to
a given sentence (support sets):
$S_{i}{=}\{s{\in} I:\text{Sim}\left(s,p_{i}\right){>}\epsilon_{i}{\wedge}
s{\neq} p_{i}\}$. Support sets are estimated for every sentence. Sentences
frequent in most support sets are selected:
$\operatorname{argmax}_{s\in\cup_{i=1}^{n}S_{i}}|\{S_{i}:s{\in} S_{i}\}|$.
This is similar to unweighted LexRank (section \ref{sub:lexrank}), except that
support sets allow a different threshold for each sentence ($\epsilon_{i}$) and
their underlying representation is directed, i.e., each sentence only recommends
its most semantically related sentences. The thresholds can be heuristically
determined. \cite{Ribeiro2011}, among others, uses a passage order heuristic
which clusters all passages into two clusters, according to their distance to
each cluster's centroid. The first and second clusters are initialized with the
first and second passages, respectively, and sentences are assigned to clusters,
one by one, according to their original order. The cluster that contains the
most similar passage to the passage associated with the support set under
construction is selected as the support set. Several metrics were tested for
defining semantic relatedness (e.g. Minkowski distance, cosine).

\section{Experiments}\label{sec:experiments}
We evaluated generic summarization by assessing its impact on binary and
multiclass music genre classification. These tasks consist of classifying songs
based on a scheme (e.g. artist, genre, or mood). Classification is deemed
important by the \MIR~community and annual conferences addressing it are held,
such as \ISMIR, which comprises \MIREX~\cite{MIREX} for comparing
state-of-the-art algorithms in a standardized setting. The best \MIREX~2015
system \cite{Wu2015}~for the ``Audio Mixed Popular Genre Classification'' task
uses \SVMs~for classifying music genre, based on spectral features.

We follow the same approach and our classification is also performed using
\SVMs~\cite{Chang2011}. Note that there are two different feature extraction
steps. The first is done by the summarizers, every time a song is summarized.
The summarizers output audio signal corresponding to the selected parts, to be
used in the second step, i.e., when doing classification, where features are
extracted from the full, segmented, and summarized datasets.

\subsection{Classification Features}\label{sub:features}
The features used by the \SVM~consist of a 38-dimensional vector per song,  a
concatenation of several statistics on features used in \cite{deLeon2013},
describing the timbral texture of a music piece. It consists of the average of
the first 20 \MFCCs~concatenated with statistics (mean and variance) of 9
spectral features: centroid, spread, skewness, kurtosis, flux, rolloff,
brightness, entropy, and flatness. These are computed over feature vectors
extracted from 50ms frames without overlap. This set of features and a smaller
set, solely composed of \MFCC~averages, were tested in the classification task.
All music genres in our dataset are timbrically different from each other,
making these sets good descriptors for classification.

\subsection{Datasets}\label{sub:datasets}
Our experimental datasets consist of a total of 1250 songs
from 5 different genres: Bass, Fado, Hip hop, Trance, and Indie Rock. Bass music is a
generic term referring to several specific styles of electronic music, such as
Dubstep, Drum and Bass, Electro, and more. Although these differ in tempo, they
share similar timbral characteristics, such as deep basslines and the
``wobble'' bass effect. Fado is a Portuguese music genre whose instrumentation
consists of stringed instruments, such as the classical and the
Portuguese guitars. Hip hop consists of drum rhythms (usually built with
samples), the use of turntables and spoken lyrics. Indie Rock
usually consists of guitar, drums, keyboard, and vocal sounds and was influenced
by punk, psychedelia, post-punk, and country. Trance is an electronic music
genre characterized by repeating melodic phrases and a musical form that builds
up and down throughout a track. Each class is represented by 250 songs from
several artists. The multiclass dataset contains all songs. Two binary datasets
were also built from this data, in order to test our hypothesis on a wider range
of classification setups: Bass vs. Fado and Bass vs. Trance, each
containing the 500 corresponding songs.

\subsection{Setup}\label{sub:setup}
10-fold cross-validation was used in all classification tasks. First, as
baselines, we performed 3 classification experiments using 30s segments, from
the beginning, middle, and end of each song. Then, we obtained another baseline
by using the whole songs. The baselines were compared with the classification
results from using 30s summaries for each parameter combination and algorithm.
We did this for both binary datasets and then for the multiclass dataset.

Applying generic summarization algorithms to music requires additional steps.
Since these algorithms operate on the discrete concepts of word and sentence,
some preprocessing must be done to map the continuous frame representation
obtained after feature extraction to a word/sentence representation. For each
song being summarized, a vocabulary is created, through clustering the frames'
feature vectors. mlpack's \cite{Curtin2013} implementation of the $K$-Means
algorithm was used for this step (we experiment with some values for $K$ and
assess their impact on the results). After clustering, a vocabulary of musical
words is obtained (each word is a frame cluster's centroid) and each frame is
assigned its own cluster centroid, effectively mapping the frame feature vectors
to vocabulary words. This transforms the real/continuous nature of each frame
(when represented by a feature vector) to a discrete nature (when represented as a
word from a vocabulary). Then, the song is segmented into fixed-size sentences
(e.g., 5-word sentences). Since every sentence contains discrete words from a
vocabulary, it is possible to represent each one as a vector of word
occurrences/frequencies (depending on the weighting scheme) which
is the exact representation used by generic summarization algorithms. Sentences
were compared using the cosine distance. The parameters of all of these
algorithms include: features, framing, vocabulary size (final number of clusters
of the $K$-Means algorithm), weighting (e.g., \emph{tf-idf}), and sentence size
(number of words per sentence).

For the multiclass dataset, we also ran experiments comparing human-oriented
summarization against generic summarization. This translates into comparing
Average Similarity summaries (for several durations) against 30-second generic
summaries, as well as comparing structural against fixed-size sentences. We also
compared the performance of generic summaries against the baselines for smaller
summary durations.

Every algorithm was implemented in C++. We used: OpenSMILE \cite{Eyben2013} for
feature extraction, Armadillo \cite{Sanderson2010} for matrix operations,
Marsyas \cite{Tzanetakis1999} for synthesizing the summaries, and the segmenter
used in \cite{Weiss2010} for structural segmentation.

Our experiments covered the following parameter values (varying between
algorithms): frame and hop size combinations of (0.25,0.125), (0.25,0.25),
(0.5,0.25), (0.5,0.5), (1,0.5) and (1,1) (in seconds); vocabulary sizes of 25,
50, and 100 (words); sentence sizes of 5, 10, and 20 (words); ``dampened''
\emph{tf-idf} (takes logarithm of \emph{tf} instead of \emph{tf} itself) and
binary weighting schemes. As summarization features, we used \MFCC~vectors of
sizes 12, 20, and 24. These features, used in several previous research efforts
on music summarization in
\cite{Chai2006,Cooper2003,Peeters2002,Peeters2003,Chu2000,Cooper2002,Glaczynski2011},
describe the timbre of an acoustic signal. We also used a concatenation of
\MFCC~vectors with the 9 spectral features enumerated in section
\ref{sub:features}. For \MMR, we tried $\lambda$ values of 0.5 and 0.7. Our
\LSA~implementation also makes use of the sentence score and the topics
cardinality selection heuristic described in section \ref{sub:lsa}.

\section{Results: Binary Tasks}\label{sec:binary-results}
First, we analyze results on the binary datasets, Bass vs. Fado and Bass vs.
Trance. The reason we chose these pairs was because we wanted to see
summarization's impact on an easy to classify dataset (Bass and Fado are
timbrically very different) and a more difficult one (Bass and Trance share many
timbrical similarities due to their electronic and dancefloor-oriented nature).
For all experiments, classifying using the 38-dimensional
features vector produced better results than using only 20 \MFCCs, so we only
present those results here. The best results are summarized in tables
\ref{tab:binary-baselines}, \ref{tab:bass-fado}, and \ref{tab:bass-trance}.

\begin{table}[htb]
\caption{Binary classification results}
\begin{subtable}{\columnwidth}
\caption{Baselines}
\begin{center}
\begin{tabular}{c|c|c}
Setup & Bass vs. Fado & Bass vs. Trance\\
\hline
Full songs & 100.0\% & 95.2\%\\
Beginning 30 s & 94.2\% & 91.4\%\\
Middle 30 s & 98.0\% & 83.6\%\\
End 30 s & 97.0\% & 89.4\%\\
\end{tabular}
\end{center}
\label{tab:binary-baselines}
\end{subtable}
~
\begin{subtable}{\columnwidth}
\caption{Bass vs. Fado summaries}
\begin{center}
\begin{tabular}{c|c|c|c|c|c}
Algorithm & Framing & Voc. & Sent. & Weight. & Accuracy\\
\hline
\GRASSHOPPER & (0.5,0.5) & 25 & 5 & binary & 100.0\%\\
LexRank & (0.5,0.5) & 25 & 10 & dampTF & 100.0\%\\
\LSA & (0.5,0.5) & 25 & 10 & binary & 100.0\%\\
\MMR & (0.5,0.5) & 25 & 10 & dampTF & 100.0\%\\
Support Sets & (0.5,0.5) & 50 & 10 & dampTF & 100.0\%\\
\end{tabular}
\end{center}
\label{tab:bass-fado}
\end{subtable}
~
\begin{subtable}{\columnwidth}
\caption{Bass vs. Trance summaries}
\begin{center}
\begin{tabular}{c|c|c|c|c|c}
Algorithm & Framing & Voc. & Sent. & Weight. & Accuracy\\
\hline
\GRASSHOPPER & (0.5,0.5) & 25 & 10 & binary & 92.2\%\\
LexRank & (0.5,0.5) & 50 & 10 & binary & 93.4\%\\
\LSA & (0.5,0.5) & 25 & 5 & binary & 93.8\%\\
\MMR & (0.5,0.5) & 25 & 5 & binary & 94.2\%\\
Support Sets & (0.5,0.5) & 25 & 10 & dampTF & 93.6\%\\
\end{tabular}
\end{center}
\label{tab:bass-trance}
\end{subtable}
\end{table}

The first thing we notice on the Bass vs. Fado task is that the middle sections
are the best continuous sections and they do a good job at distinguishing Fado
from other genres. Accuracy dropped just 2 \pp~against using full songs.
However, the beginning sections' accuracy dropped by 5.8\pp. All summarization
algorithms fully recovered the accuracy lost by any continuous sections against
using full songs, achieving the 100\% full songs baseline. In this case,
summarization helps classification in an already easy task. The $\lambda$ value
in \MMR's setup was 0.7 and the passage order heuristic using the cosine
similarity was used for calculating the support sets.

In the Bass vs. Trance task, the middle sections do a very poor job at
describing and distinguishing these genres -- they actually perform worse than
the beginning or end sections. Actually, the worst sections in the Bass vs.
Fado task were the best in this one and vice-versa. This means that choosing a
continuous segment to extract features for classification purposes cannot be
assumed to work equally well for every genre and dataset. All summarization
algorithms, while not reaching the same performance as when using full songs,
succeeded in improving classification performance against the continuous
30-second baselines.
In this case, summarization is helping classification in a more difficult task.
Again, \MMR's $\lambda$ value was set to 0.7 and the passage order heuristic
using the cosine similarity was used for calculating the support sets.

\section{Results: Multiclass Tasks}\label{sec:multiclass-results}
Since we are extrinsically evaluating summarization, analyzing its impact on
music classification must go beyond simply comparing final classification
accuracy for each scenario (as was done for binary classification). Here, we
also look at the confusion matrices obtained from the classification scenarios,
so that we can carefully look at the data (in this case, listen to the data) to
understand what is happening when summarizing music this way and why it is
improving the classification task's performance. Since our dataset consists of
250 songs per class, each confusion matrix row must sum to 250. Classes are
identically sorted both in rows and columns, which means the ideal case is where
we have a diagonal confusion matrix (all zeros, except for the diagonal
elements, which should all be 250). Class name initials are shown to the left of
the matrix and individual class accuracies are shown to the right.

\subsection{Full songs}
First, we look at the confusion matrix resulting from classifying full songs
(table \ref{tab:full}). We can see that Fado, although there is some confusion
between it and Indie Rock, is the most distinguishable genre within this group
of genres which makes sense since timbrically it is very different from every
other genre present in the dataset. Trance and Bass also achieve accuracies over
90\%, although they also share some confusion which is explained by the fact
that they both are Electronic music styles, thus sharing many timbral characteristics
derived from the virtual instruments used to produce them. The classifier
performs worse when classifying Hip hop and Indie Rock, achieving accuracies
around 84\% and confusing both genres in approximately 10\% of the tracks. This
can also be explained by the fact that both of those genres have strong vocals
presence (in contrast with Bass and Trance). Although Fado also has an important
vocal component, its instrumentation is very different from Hip hop and Indie
Rock explaining why Fado did not get confused as much as they were with
each other. The overall accuracy of this classification scenario is 89.84\%.

We can think of these accuracies as how well these classification features (and
\SVM) can perform on these genres, given all the possible information about the
tracks. Intuitively, removing information by, for instance, only extracting
features from the beginning 30 seconds of the songs, will worsen the performance
of the classifier because it will have incomplete data about each song, and
thus, also incomplete data for modeling each class. Tables \ref{tab:beginning},
\ref{tab:middle}, and \ref{tab:end} show that to be true when using such a blind
approach to summarize music (since extracting 30-second contiguous segments can
also be interpreted as a naive summarization method). This process of extracting
features from a dataset of segments is what is usually done when classifying
music, since processing 30 seconds instead of the whole song saves processing
time.

\begin{table}[htb]
\caption{Full songs (89.8\%)}
\begin{center}
\begin{tabular}{c|ccccc|c}
B & 226 & 0 & 6 & 1 & 17 & 90.4\%\\
F & 0 & 240 & 1 & 9 & 0 & 96.0\%\\
H & 8 & 2 & 209 & 27 & 4 & 83.6\%\\
I & 0 & 14 & 24 & 212 & 0 & 84.8\%\\
T & 10 & 0 & 4 & 0 & 236 & 94.4\%\\
\end{tabular}
\end{center}
\label{tab:full}
\end{table}

\subsection{Baseline segments}
Table \ref{tab:beginning} shows classification results when using only the 30
seconds from the beginning of the songs. Table \ref{tab:beginning-full} shows
the comparison of the beginning sections against full songs. The classification
accuracy is 77.52\%, i.e., a 12.32\pp~drop when compared to using full songs.
Bass accuracy dropped 19.6\pp, due to increased confusion with both Hip hop and
Indie Rock. Trance was also more confused with Indie Rock. This is easily
explained by the fact that the first 30 seconds of most Bass or Trance songs
correspond to the \emph{intro} part. These \emph{intros} are lower energy parts
which may contain a relatively strong vocal presence and much fewer
instrumentation than other more characteristic parts of the genres.
These \emph{intros} are much more similar to Hip hop and Indie Rock
\emph{intros} than when considering the whole songs, explaining why the
classifier is confusing these classes more in this scenario. Thus, taking the
beginning of the songs for classification is, in general, not a good
summarization strategy.

\small

\setlength{\tabcolsep}{1.5mm}
\newlength{\x}\settowidth{\x}{444}
\newcommand{\xx}[1]{\hbox to .9\x{\hfil#1}}

\begin{table*}[htb]
\caption{Baseline confusion matrices}
\centering
\begin{subtable}{.3\textwidth}
\caption{Beginning sections (77.5\%)}
\begin{center}
\begin{tabular}{c|rrrrr|r}
B & \xx{177} & \xx{6} & \xx{33} & \xx{29} & \xx{5} & 70.8\%\\
F & 8 & 221 & 2 & 19 & 0 & 88.4\%\\
H & 32 & 1 & 179 & 31 & 7 & 71.6\%\\
I & 27 & 15 & 22 & 182 & 4 & 72.8\%\\
T & 14 & 0 & 12 & 14 & 210 & 84.0\%\\
\end{tabular}
\end{center}
\label{tab:beginning}
\end{subtable}
~
\begin{subtable}{.3\textwidth}
\caption{Middle sections (81.4\%)}
\begin{center}
\begin{tabular}{c|rrrrr|r}
B & \xx{184} & \xx{0} & \xx{18} & \xx{4} & \xx{44} & 73.6\%\\
F & 2 & 233 & 2 & 13 & 0 & 93.2\%\\
H & 14 & 4 & 206 & 18 & 8 & 82.4\%\\
I & 2 & 10 & 12 & 209 & 17 & 83.6\%\\
T & 34 & 3 & 8 & 20 & 185 & 74.0\%\\
\end{tabular}
\end{center}
\label{tab:middle}
\end{subtable}
~
\begin{subtable}{.3\textwidth}
\caption{End sections (76.8\%)}
\begin{center}
\begin{tabular}{c|rrrrr|r}
B & \xx{178} & \xx{3} & \xx{32} & \xx{19} & \xx{18} & 71.2\%\\
F & 4 & 223 & 5 & 17 & 1 & 89.2\%\\
H & 32 & 8 & 175 & 31 & 4 & 70.0\%\\
I & 18 & 14 & 38 & 175 & 5 & 70.0\%\\
T & 23 & 1 & 11 & 6 & 209 & 83.6\%\\
\end{tabular}
\end{center}
\label{tab:end}
\end{subtable}
~\\
~\\
~\\
\begin{subtable}{.3\textwidth}
\caption{Beginning vs Full (-12.3\%)}
\begin{center}
\begin{tabular}{c|rrrrr|r}
B & \xx{-49} & \xx{6} & \xx{27} & \xx{28} & \xx{-12} & -19.6\%\\
F & 8 & -19 & 1 & 10 & 0 & -7.6\%\\
H & 24 & -1 & -30 & 4 & 3 & -12.0\%\\
I & 27 & 1 & -2 & -30 & 4 & -12.0\%\\
T & 4 & -10 & 8 & 14 & -26 & -10.4\%\\
\end{tabular}
\end{center}
\label{tab:beginning-full}
\end{subtable}
~
\begin{subtable}{.3\textwidth}
\caption{Middle vs Full (-8.5\%)}
\begin{center}
\begin{tabular}{c|rrrrr|r}
B & \xx{-42} & \xx{0} & \xx{12} & \xx{3} & \xx{27} & -16.8\%\\
F & 2 & -7 & 1 & 4 & 0 & -2.8\%\\
H & 6 & 2 & -3 & -9 & 4 & -1.2\%\\
I & 2 & -4 & -12 & -3 & 17 & -1.2\%\\
T & 24 & -7 & 4 & 20 & -51 & -20.4\%\\
\end{tabular}
\end{center}
\label{tab:middle-full}
\end{subtable}
~
\begin{subtable}{.3\textwidth}
\caption{End vs Full (-13.0\%)}
\begin{center}
\begin{tabular}{c|rrrrr|r}
B & \xx{-48} & \xx{3} & \xx{26} & \xx{18} & \xx{1} & -19.2\%\\
F & 4 & -17 & 4 & 8 & 1 & -6.8\%\\
H & 24 & 6 & -34 & 4 & 0 & -13.6\%\\
I & 18 & 0 & 14 & -37 & 5 & -14.8\%\\
T & 13 & -9 & 7 & 6 & -27 & -10.8\%\\
\end{tabular}
\end{center}
\label{tab:end-full}
\end{subtable}
\label{tab:multiclass-baselines}
\end{table*}

\normalsize

Tables \ref{tab:middle} and \ref{tab:middle-full} show classification results
when using the middle 30 seconds of the songs and the comparison of those
segments against full songs, respectively. The overall accuracy was 81.36\%,
i.e., an 8.48\pp~drop against the full songs baseline. This time, both
Bass and Trance accuracies dropped 16.8\pp~and 20.4\pp, respectively, getting
confused with each other by the classifier. Having listened to the tracks that
got confused this way, the conclusion is as expected: these middle segments
correspond to what is called a \emph{breakdown} section of the songs. These
sections correspond to lower energy segments (though not as low as an
\emph{intro}) of the tracks which, again, are not the most characteristic parts
of both these genres and, in the particular case of Bass vs. Trance, they are
timbrically very similar due to their Electronic nature. A human listener would,
probably, also be unable to distinguish between these two genres if listening
only to these segments. Although, for 3 of the 5 genres, classification
performance did not drop pronouncedly, it did so for 2 of them, which means
that, in general, taking the middle sections of the songs for classification is
also not a good segment selection strategy.

Tables \ref{tab:end} and \ref{tab:end-full} show classification results when
using the last 30 seconds of the songs and the comparison of those segments
versus full songs, respectively. The end sections obtained an accuracy of
76.8\%, i.e., a 13.04\pp~decrease when compared against full songs. Again, Bass
was mainly misclassified as Hip hop and Indie Rock, and Trance was mainly
misclassified as Bass. This is mostly due to the fact that the last 30 seconds
correspond to the \emph{outro} section of the songs which shares many
similarities with the \emph{intro} section. When considering Trance and Bass,
the \emph{outro} also shares characteristics with the \emph{breakdown} sections.
The fade repeat effect present in many songs' endings also increases this
confusion. This means that taking the last 30 seconds of a song is also not a
good segment selection strategy.

\subsection{Baseline Assessment}
Although, from the above experiments, it seems that taking the middle sections
of the songs is better than taking the beginning or end, it is still not
good enough, at least, not for all of the considered genres. The features used
by the classifier are statistics (means and variances) of features extracted
along the whole signal. Those features perform well when taking the whole signal as
input, which means that, in order to obtain a similar performance, those
statistics should be similar. That cannot be guaranteed when taking
30-second continuous clips because those 30 seconds may happen to belong to a
single (and not distinctive enough) structural part of the song (such as
\emph{intro}, \emph{breakdown}, and \emph{outro}). If that is the case, then
there is not sufficient diversity in the segment/summary to accurately represent
the whole song. Moreover, some music genres can only be accurately distinguished
by some of those structural parts: the best examples in this dataset are the
Bass and Trance classes, which are much more accurately distinguished and
represented by their \emph{drop} sections. Therefore, we need to make better
choices regarding what parts of the song should be included in the 30-second
summaries to be classified.

\subsection{\GRASSHOPPER}
Generic summarization algorithms define and detect relevance and diversity of
the input signal, satisfying our need for a more informed way of selecting the
most important parts to fit in 30-second summaries. The following tables show
results demonstrating this claim. Tables \ref{tab:grasshopper} and
\ref{tab:grasshopper-middle} show classification results when using summaries
extracted by \GRASSHOPPER. The specific parameter values used in this experiment
were: (0.5,0.5) seconds framing, 25-word vocabulary, 10-word sentences, and
binary weighting. The overall accuracy was 88.16\%. As can be seen,
\GRASSHOPPER~recovered most of what was lost by the middle sections, in terms of
classification accuracy for each class. Since the middle sections performed so
badly when distinguishing Bass and Trance, naturally, these summaries improved
accuracies mostly for both these classes, with 14.0\pp~and 14.4\pp~increases,
respectively. When listening to some of these summaries, the diversity included
in them is clear: the algorithm is selecting sentences from several different
structural parts of the songs. An overall improvement of 6.80\pp~was obtained
this way. Note that, remarkably, these summaries did a better job than full
songs at classifying Hip hop by 2.0\pp. This means that, for some tasks, well
summarized data can be even more discriminative of a topic (genre) than the
original full data.

\begin{table}[htb]
\caption{\GRASSHOPPER}
\begin{subtable}{\columnwidth}
\caption{Summaries (88.2\%)}
\begin{center}
\begin{tabular}{c|rrrrr|r}
B & 219 & 0 & 7 & 5 & 19 & 87.6\%\\
F & 0 & 235 & 4 & 11 & 0 & 94.0\%\\
H & 8 & 6 & 214 & 18 & 4 & 85.6\%\\
I & 6 & 14 & 17 & 213 & 0 & 85.2\%\\
T & 23 & 0 & 5 & 1 & 221 & 88.4\%\\
\end{tabular}
\end{center}
\label{tab:grasshopper}
\end{subtable}
~\\
~\\
\begin{subtable}{\columnwidth}
\caption{Summaries vs. Middle sections (6.8\%)}
\begin{center}
\begin{tabular}{c|rrrrr|r}
B & 35 & 0 & -11 & 1 & -25 & 14.0\%\\
F & -2 & 2 & 2 & -2 & 0 & 0.8\%\\
H & -6 & 2 & 8 & 0 & -4 & 3.2\%\\
I & 4 & 4 & 5 & 4 & -17 & 1.6\%\\
T & -11 & -3 & -3 & -19 & 36 & 14.4\%\\
\end{tabular}
\end{center}
\label{tab:grasshopper-middle}
\end{subtable}
\end{table}

\subsection{LexRank}
Tables \ref{tab:lexrank} and \ref{tab:lexrank-middle} present the LexRank
confusion matrix and its difference against the middle sections. The parameter
values in this experiment were: (0.5,0.5) seconds framing, 25-word vocabulary,
5-word sentences and dampened \emph{tf-idf} weighting. The overall accuracy was
88.40\%. LexRank also greatly improved classification accuracy, when compared
against the middle sections (7.04\pp~overall), namely, for Bass and Trance,
with 15.6\pp~and 15.2\pp~increases, respectively. LexRank is clearly selecting
diverse parts to include in the 30-second summaries, as we were able to conclude
when listening to them. It is also interesting that the classifier performed
better than with full songs, individually, for another class: Indie Rock's
accuracy increased 1.6\pp.

\begin{table}[htb]
\caption{LexRank}
\begin{subtable}{\columnwidth}
\caption{Summaries (88.4\%)}
\begin{center}
\begin{tabular}{c|rrrrr|r}
B & 223 & 0 & 5 & 2 & 20 & 89.2\%\\
F & 0 & 238 & 1 & 11 & 0 & 95.2\%\\
H & 9 & 6 & 205 & 24 & 6 & 82.0\%\\
I & 1 & 12 & 20 & 216 & 1 & 86.4\%\\
T & 20 & 0 & 4 & 3 & 223 & 89.2\%\\
\end{tabular}
\end{center}
\label{tab:lexrank}
\end{subtable}
~\\
~\\
\begin{subtable}{\columnwidth}
\caption{Summaries vs. Middle sections (7.0\%)}
\begin{center}
\begin{tabular}{c|rrrrr|r}
B & 39 & 0 & -13 & -2 & -24 & 15.6\%\\
F & -2 & 5 & -1 & -2 & 0 & 2.0\%\\
H & -5 & 2 & -1 & 6 & -2 & -0.8\%\\
I & -1 & 2 & 8 & 7 & -16 & 2.8\%\\
T & -14 & -3 & -4 & -17 & 38 & 15.2\%\\
\end{tabular}
\end{center}
\label{tab:lexrank-middle}
\end{subtable}
\end{table}

\subsection{\LSA}
Tables \ref{tab:lsa} and \ref{tab:lsa-middle} show the \LSA~confusion matrix of
and the corresponding difference against the middle sections. The following
parameter combination was used: (0.5,0.5) seconds framing, 25-word vocabulary,
10-word sentences, and binary weighting. Note that using a term frequency-based
weighting on \LSA, when applied to music, markedly worsens its performance. This
is because noisy sentences in the songs tend to get a very high score on some
latent topic, causing \LSA~to include them in the summaries. Moreover, when also
considering inverse document frequency, the results are even worse, because
those noisy terms usually appear in very few sentences. That is highly
undesirable, since those sections do a very bad job at describing that song in
any aspect. Using a binary weighting scheme alleviates that problem because all
those noisy frames will get clustered into very few clusters/terms and only that
term's presence (instead of frequency) gets counted into the sentences' vector
representation. The overall accuracy for this combination was 88.32\%, an
improvement of 6.98\pp~against the middle sections.
Bass and Trance were also the genres which benefited the most from this
summarization, with accuracy increases of 12.8\pp~and 14.8\pp, respectively,
which can also be explained by the diversity present in the summaries.
Indie Rock's individual accuracy improved, once again, against full songs, with
an improvement of 2.8\pp.

\begin{table}[htb]
\caption{\LSA}
\begin{subtable}{\columnwidth}
\caption{Summaries (88.3\%)}
\begin{center}
\begin{tabular}{c|rrrrr|r}
B & 216 & 0 & 13 & 2 & 19 & 86.4\%\\
F & 0 & 241 & 2 & 7 & 0 & 96.4\%\\
H & 16 & 2 & 206 & 24 & 2 & 82.4\%\\
I & 3 & 14 & 14 & 219 & 0 & 87.6\%\\
T & 20 & 0 & 7 & 1 & 222 & 88.0\%\\
\end{tabular}
\end{center}
\label{tab:lsa}
\end{subtable}
~\\
~\\
\begin{subtable}{\columnwidth}
\caption{Summaries vs. Middle sections (7.0\%)}
\begin{center}
\begin{tabular}{c|rrrrr|r}
B & 32 & 0 & -5 & -2 & -25 & 12.8\%\\
F & -2 & 8 & 0 & -6 & 0 & 3.2\%\\
H & 2 & -2 & 0 & 6 & -6 & 0.0\%\\
I & 1 & 4 & 2 & 10 & -17 & 4.0\%\\
T & -14 & -3 & -1 & -19 & 37 & 14.8\%\\
\end{tabular}
\end{center}
\label{tab:lsa-middle}
\end{subtable}
\end{table}

\subsection{\MMR}
Tables \ref{tab:mmr} and \ref{tab:mmr-middle} represent the confusion matrix for
an \MMR~summarization setup and its difference against the middle sections.
(0.5,0.5) seconds framing was used, along with a 50-word vocabulary, 10-word
sentences, 0.7 $\lambda$ value, and dampened \emph{tf-idf} weighting. Note that,
even though every other parameter setup (for the other algorithms) shown here
uses 20 \MFCCs~as features, this one uses those same \MFCCs~concatenated with
the 9 spectral features also used for classification (described in section
\ref{sub:features}). This is because \MMR, unlike every other summarization
algorithm, performed better using this set (instead of only using \MFCCs~as
features). The overall accuracy was 88.80\%, corresponding to an improvement of
7.44\pp~over the middle sections. Bass and Trance benefited the most from the
summarization process, in classification performance, achieving improvements of
14.8\pp~and 16.4\pp, respectively. This is also explained by the diversity
produced by the summarizer.

\begin{table}[htb]
\caption{\MMR}
\begin{subtable}{\columnwidth}
\caption{Summaries (88.8\%)}
\begin{center}
\begin{tabular}{c|rrrrr|r}
B & 221 & 0 & 9 & 2 & 18 & 88.4\%\\
F & 0 & 242 & 2 & 6 & 0 & 96.8\%\\
H & 11 & 1 & 210 & 24 & 4 & 84.0\%\\
I & 1 & 13 & 25 & 211 & 0 & 84.4\%\\
T & 20 & 0 & 4 & 0 & 226 & 90.4\%\\
\end{tabular}
\end{center}
\label{tab:mmr}
\end{subtable}
~\\
~\\
\begin{subtable}{\columnwidth}
\caption{Summaries vs. Middle sections (7.4\%)}
\begin{center}
\begin{tabular}{c|rrrrr|r}
B & 37 & 0 & -9 & -2 & -26 & 14.8\%\\
F & -2 & 9 & 0 & -7 & 0 & 3.6\%\\
H & -3 & -3 & 4 & 6 & -4 & 2.4\%\\
I & -1 & 3 & 13 & 2 & -17 & 0.8\%\\
T & -14 & -3 & -4 & -20 & 41 & 16.4\%\\
\end{tabular}
\end{center}
\label{tab:mmr-middle}
\end{subtable}
\end{table}

\subsection{Support Sets}
Tables \ref{tab:support-sets} and \ref{tab:support-sets-middle} show results
obtained when classifying the dataset using summaries extracted by the Support
Sets-based algorithm. The specific parameter setup of this experiment was:
(0.5,0.5) seconds framing, 25-word vocabulary, 10-word sentences, dampened
\emph{tf-idf} weighting, and the passage order-based heuristic for creating the
support sets \cite{Ribeiro2011} using the cosine similarity. The overall
accuracy was 88.80\%. Again, summarization recovered most of what was lost by
the middle sections in terms of classification accuracy for each individual
class, greatly influencing Bass and Trance, with 10.8\pp~and 16.8
\pp~increases, respectively. Listening to some of these summaries, we confirmed
the diversity included in them that was clearly lacking in the middle sections.
An overall improvement of 7.44\pp~was obtained this way. Remarkably, there were
also improvements against full songs, namely, a 4.8\pp~improvement in Indie
Rock.

\begin{table}[htb]
\caption{Support Sets}
\begin{subtable}{\columnwidth}
\caption{Summaries (88.8\%)}
\begin{center}
\begin{tabular}{c|rrrrr|r}
B & 211 & 0 & 14 & 3 & 22 & 84.4\%\\
F & 0 & 237 & 0 & 13 & 0 & 94.8\%\\
H & 7 & 4 & 211 & 22 & 6 & 84.4\%\\
I & 1 & 8 & 16 & 224 & 1 & 89.6\%\\
T & 13 & 0 & 6 & 4 & 227 & 90.8\%\\
\end{tabular}
\end{center}
\label{tab:support-sets}
\end{subtable}
~\\
~\\
\begin{subtable}{\columnwidth}
\caption{Summaries vs. Middle sections (7.4\%)}
\begin{center}
\begin{tabular}{c|rrrrr|r}
B & 27 & 0 & -4 & -1 & -22 & 10.8\%\\
F & -2 & 4 & -2 & 0 & 0 & 1.6\%\\
H & -7 & 0 & 5 & 4 & -2 & 2.0\%\\
I & -1 & -2 & 4 & 15 & -16 & 6.0\%\\
T & -21 & -3 & -2 & -16 & 42 & 16.8\%\\
\end{tabular}
\end{center}
\label{tab:support-sets-middle}
\end{subtable}
\end{table}

\subsection{Summary size experiments}
\begin{figure}[htb]
\begin{center}
\includegraphics[keepaspectratio=true,width=\linewidth]{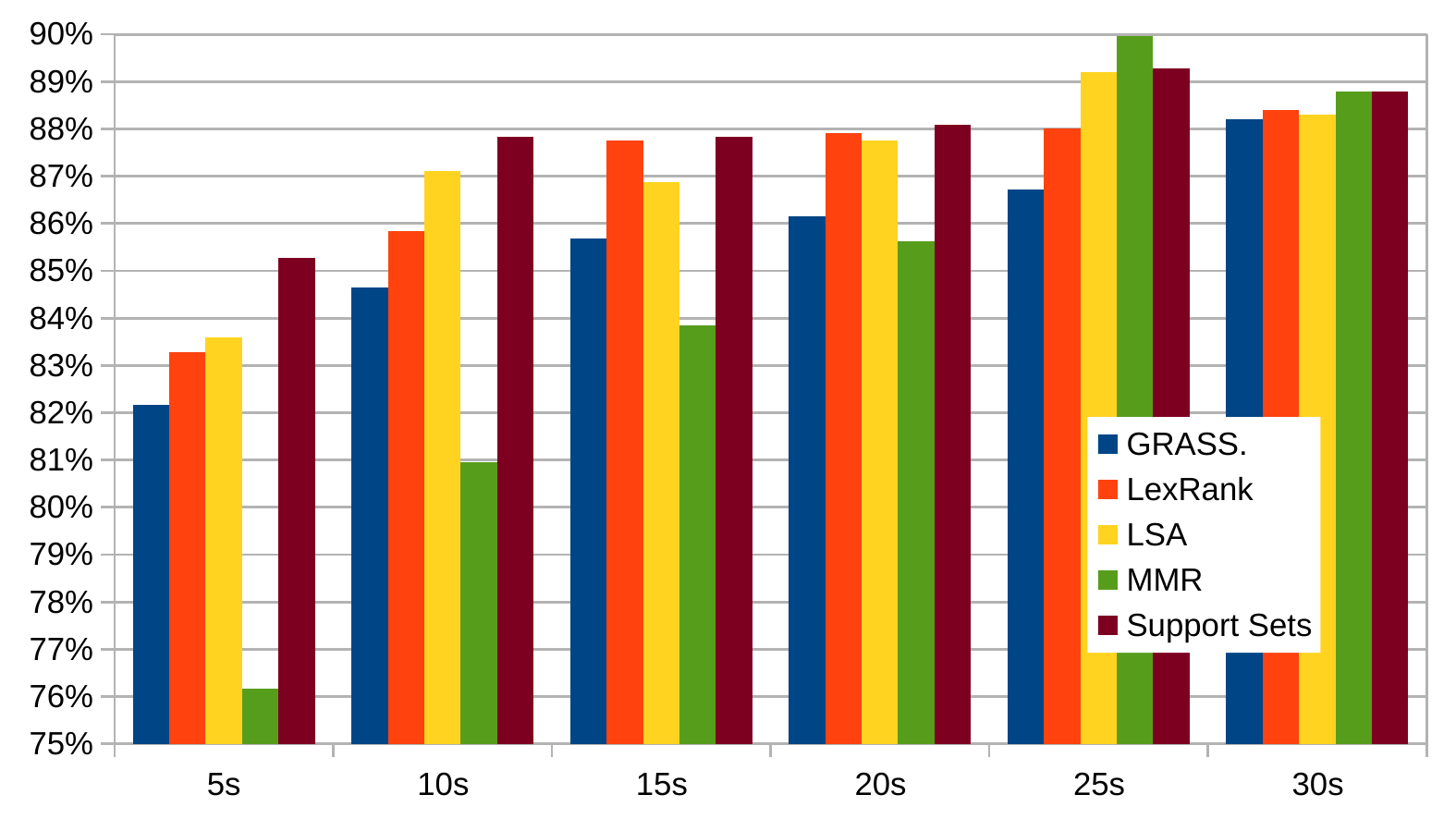}
\end{center}
\caption{Accuracy (\%) vs summary size (s). Baselines accuracies are
77.5\%, 81,4\%, and 76.8\% for the beginning, middle, and end sections,
respectively. Full songs achieve 89.8\%.}
\label{fig:summary-sizes}
\end{figure}

To better evaluate the robustness of these methods, we ran experiments using
decreasing summary sizes. For these experiments, no search for optimal parameter
combinations was done: we used the ones that maximized classification accuracy
for 30-second summaries. These are not necessarily the best parameters for
smaller summary sizes but allow using the 30-second summaries as baselines. We
ran these experiments for summary sizes of 5 to 25 seconds and report the
results in table \ref{tab:summary-sizes} and Figure \ref{fig:summary-sizes}.

\begin{table}[htb]
\caption{Summary size experiments}
\begin{center}
\begin{tabular}{r|rrrrr}
 & GRASSH. & LexRank & \LSA & \MMR & Support Sets\\
\hline
5 s & 82.16\% & 83.28\% & 83.60\% & 76.16\% & 85.28\%\\
10 s & 84.64\% & 85.84\% & 87.12\% & 80.96\% & 87.84\%\\
15 s & 85.68\% & 87.76\% & 86.88\% & 83.84\% & 87.84\%\\
20 s & 86.16\% & 87.92\% & 87.76\% & 85.36\% & 88.08\%\\
25 s & 86.72\% & 88.00\% & 89.20\% & 86.96\% & 89.28\%\\
\end{tabular}
\end{center}
\label{tab:summary-sizes}
\end{table}

Considering classification accuracy, every algorithm, except for \MMR,
outperforms the best 30-second baseline with just 5-second summaries. \LSA~and
Support Sets, in particular, surpass the 87\% accuracy mark using just 10-second
summaries. Note that these experiments were not fine tuned.

\subsection{Average Similarity}
\begin{figure}[htb]
\begin{center}
\includegraphics[keepaspectratio=true,width=\linewidth]{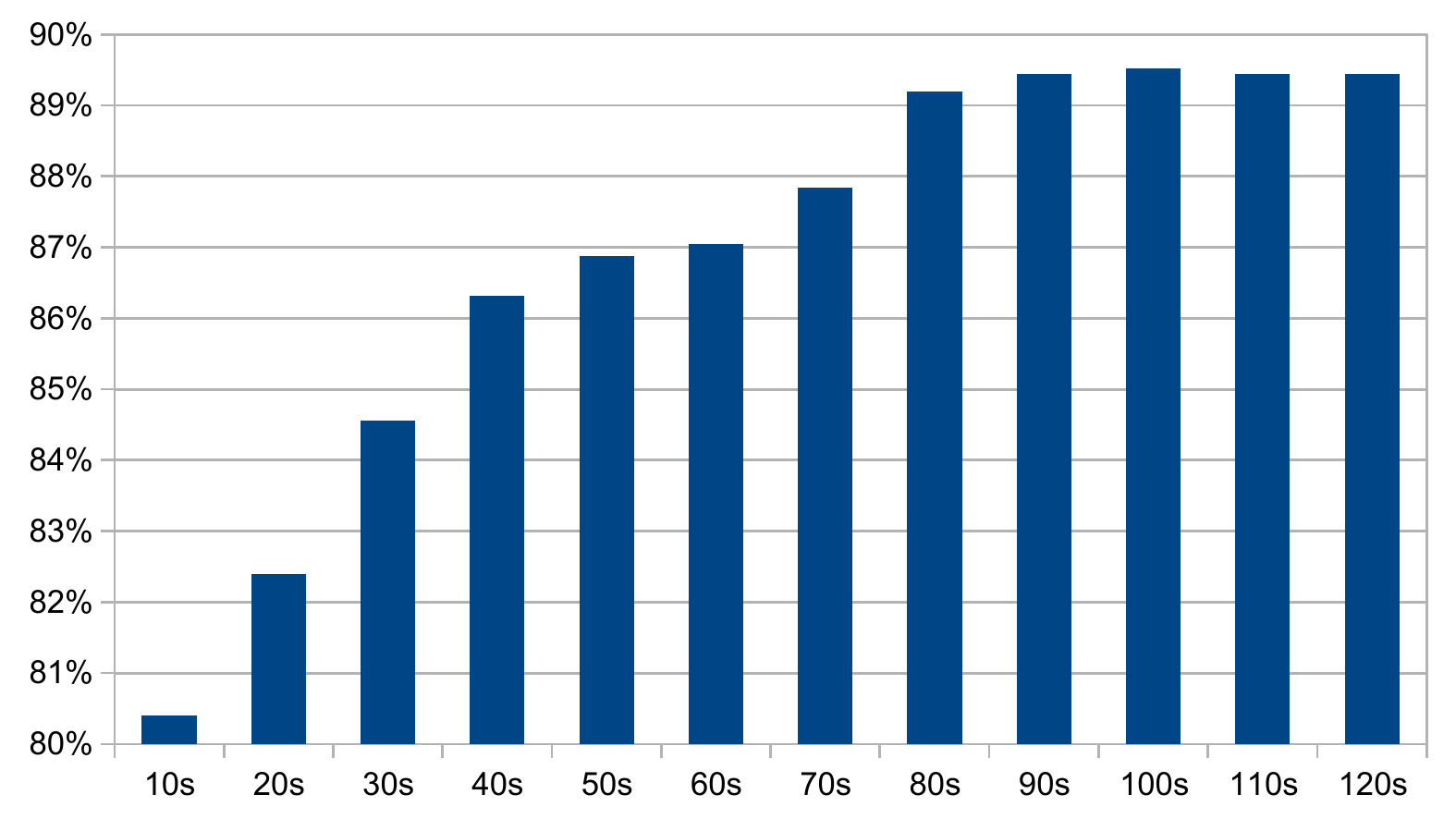}
\end{center}
\caption{Average Similarity accuracy (\%) vs summary size (s).}
\label{fig:avg-sim}
\end{figure}

To obtain a human-oriented baseline, we summarized the dataset with Average
Similarity (section \ref{sec:music-summarization}). This can be seen as an
informed human-relevant way of selecting the best starting position of a
contiguous segment. The parameter values used in this experiment were:
(0.5,0.5) seconds framing, and the first 20 MFCCs as features. Since this
algorithm does not explicitly account for diversity, we summarized using several
durations, to assess the required summary length for this type of summarization
to achieve the same classification performance of full songs or generic
summarization. We report these results in Table \ref{tab:avg-sim} and Figure
\ref{fig:avg-sim}.

\begin{table}[htb]
\caption{Average Similarity summaries}
\begin{center}
\begin{tabular}{c|ccccccccc}
Dur. (s) & 10 & 20 & 30 & 40 & 50 & 60\\
Acc. (\%) & 80.4 & 82.4 & 84.56 & 86.32 & 86.88 & 87.04\\
Dur. (s) & 70 & 80 & 90 & 100 & 110 & 120\\
Acc. (\%) & 87.84 & 89.2 & 89.44 & 89.52 & 89.44 & 89.44\\
\end{tabular}
\end{center}
\label{tab:avg-sim}
\end{table}

We can see that this type of summarization reaches the performance of generic
summaries (30 seconds) and full songs when the summary duration reaches 80
seconds (89.2\% accuracy). This means that, for a human-oriented
summary to be as descriptive and discriminative as a generic summary, an
additional 50 seconds (2.67 times the length of the original) are needed. Even
though the starting point of this contiguous summary is carefully selected by this
algorithm, it still lacks diversity because of its contiguous nature, hindering
classification accuracy for this summarizer. Naturally, by extending summary
duration, summaries include more diverse information, eventually
achieving the accuracy of full songs.

\subsection{Structurally segmented sentences}
Another form of human-oriented summarization is achieved by using generic
summarization operating on structurally segmented sentences, done according to
what humans might consider to be structurally meaningful segments. After
structural segmentation, we fed each of the 5 generic algorithms with the
resulting sentences instead of fixed-size ones and truncated the summary at 30
seconds, when necessary. The parameterization used for these experiments was the
one that yielded the best results in the previous experiments for each
algorithm.

The accuracy results for \GRASSHOPPER, LexRank, \LSA, \MMR, and Support Sets
were, respectively, 82.64\%, 83.76\%, 81.84\%, 82.40\%, and 83.84\%. Even though
structurally segmented sentences slightly improve performance, when considering
classification accuracy, they are still outperformed by fixed-size segmentation.
The best algorithm can only achieve 83.84\% accuracy. This is because these
sentences are much longer, therefore harming diversity in summaries.
Furthermore, important content in structural sentences can always be
extracted when using smaller fixed-size sentences. Thus, using smaller
sentences, prevents the selection of redundant content.

\section{Discussion}\label{sec:discussion}
We ran the Wilcoxon signed-ranked test on all of the confusion matrices
presented above against the full songs scenario. The continuous sections'
p-values were $3.104\times10^{-4}$, $3.628\times10^{-3}$, and
$2.858\times10^{-5}$ for the 30-second beginning, middle, and end sections of
the songs, respectively, which means that they differ markedly from using full
songs (as can also be seen by the accuracy drops they cause). The summaries,
however, were very close to full songs, in terms of accuracy. The p-values for
\GRASSHOPPER, LexRank, \LSA, \MMR, and Support Sets were $0.10$, $0.09$, $0.16$,
$0.20$, and $0.22$, respectively. Thus, statistically speaking, using any of
these 30-second summaries does not significantly differ from using full songs
for classification (considering 95\% confidence intervals). Furthermore, the
p-values for 20-second \LSA~summaries and for 10-second Support Sets summaries
were $0.06$ and $0.08$, respectively, with the remaining p-values of increasing
summary sizes also being superior to $0.05$. Thus, statistically speaking,
generic summarization (in some cases) does not significantly differ from using
full songs for classification, for summaries as short as 10 seconds (considering
a 95\% confidence interval). This is noteworthy, considering that the average
song duration in this dataset is 283 seconds, which means that we achieve
similar levels of classification performance using around 3.5\% of the data.
Human-oriented summarization is able to achieve these performance levels, but
only at 50-second summaries and with a p-value of $0.055$, barely over the
$0.05$ threshold. However, the 60-second summaries produced by this algorithm
cannot reach that threshold. Only at 80 seconds is a comfortable p-value
($0.38$) for the 95\% confidence interval attained.



Although every algorithm creates summaries in a different way, they all tend to
include relevant and diverse sentences. This compensates their reduced lengths
(up to 30 seconds of audio) allowing those clips to be representative of the
whole musical pieces, from an automatic consumption view, as demonstrated by our
experiments. Moreover, choosing the best 30-second contiguous segments is highly
dependent on the genres in the dataset and tasks it will be used for, which is
another reason for preferring summaries over those segments. The more varied the
dataset, the less likely a fixed continuous section extraction method is to
produce representative enough clips. Bass and Trance were the most influenced
genres, by summarization, in these experiments. These are styles with very well
defined structural borders, and a very descriptive structural element -- the
\emph{drop}. The lack of that same element in a segment markedly hinders
classification performance, suggesting that any genre with similar
characteristics may also benefit from this type of summarization. It is also
worth restating that Hip hop and Indie Rock were very positively influenced by
summarization, regarding classification performance improvements over
using full songs. This shows that, sometimes, classification on summarized music
can even outperform using the whole data from the original signal. We also
demonstrated that generic summarization using fixed-size sentences, that is,
summarization not specifically oriented towards human consumption greatly
outperforms human-oriented summarization approaches for the classification task.

Summarizing music prior to the classification task also takes time, but we do
not claim it is worth doing it every time we are about do perform a \MIR~task.
The idea is to compute summarized datasets offline for future use in any task
that can benefit from them (e.g., music classification). Currently, sharing
music datasets for \MIR~research purposes is very limited in many aspects, due
to copyright issues. Usually, datasets are shared through features extracted
from (30-second) continuous clips. That practice has drawbacks, such as: those
30 seconds may not contain the most relevant information and may even be highly
redundant; and the features provided may not be the ones a researcher needs for
his/her experiments. Summarizing datasets this way also helps avoiding copyright
issues (because summaries are not created in a way enjoyable by humans) and
still provide researchers with the most descriptive parts (according to each
summarizer) of the signal itself, so that many different kinds of features can
be extracted from them.

\section{Conclusions and Future Work}\label{sec:conclusions}
We showed that generic summarization algorithms perform well when summarizing
music datasets about to be classified. The resulting summaries are remarkably
more descriptive of the whole songs than their continuous segments (of the same
duration) counterparts. Sometimes, these summaries are even more discriminative
than the full songs. We also presented an argument stating some advantages in
sharing summarized datasets within the \MIR~community.

An interesting research direction would be to automatically determine the best
vocabulary size for each song. Testing summarization's performance on different
classification tasks (e.g., with more classes) is also necessary to further
strengthen our conclusions. More comparisons with non-contiguous human-oriented
summaries should also be done. More experimenting should be done in other
\MIR~tasks that also make use of only a portion of the whole signal.

\bibliographystyle{IEEEtran}

\begin{thebibliography}{10}
\providecommand{\url}[1]{#1}
\csname url@samestyle\endcsname
\providecommand{\newblock}{\relax}
\providecommand{\bibinfo}[2]{#2}
\providecommand{\BIBentrySTDinterwordspacing}{\spaceskip=0pt\relax}
\providecommand{\BIBentryALTinterwordstretchfactor}{4}
\providecommand{\BIBentryALTinterwordspacing}{\spaceskip=\fontdimen2\font plus
\BIBentryALTinterwordstretchfactor\fontdimen3\font minus
  \fontdimen4\font\relax}
\providecommand{\BIBforeignlanguage}[2]{{%
\expandafter\ifx\csname l@#1\endcsname\relax
\typeout{** WARNING: IEEEtran.bst: No hyphenation pattern has been}%
\typeout{** loaded for the language `#1'. Using the pattern for}%
\typeout{** the default language instead.}%
\else
\language=\csname l@#1\endcsname
\fi
#2}}
\providecommand{\BIBdecl}{\relax}
\BIBdecl

\bibitem{Chai2006}
W.~Chai, ``{Semantic Segmentation and Summarization of Music: Methods Based on
  Tonality and Recurrent Structure},'' \emph{{IEEE Signal Processing
  Magazine}}, vol.~23, no.~2, pp. 124--132, 2006.

\bibitem{Cooper2003}
M.~Cooper and J.~Foote, ``{Summarizing Popular Music via Structural Similarity
  Analysis},'' in \emph{{Proc. of the IEEE Workshop on Applications of Signal
  Processing to Audio and Acoustics}}, 2003, pp. 127--130.

\bibitem{Peeters2002}
G.~Peeters, A.~{La Burthe}, and X.~Rodet, ``{Toward Automatic Music Audio
  Summary Generation from Signal Analysis},'' in \emph{{Proc. of the 3rd ISMIR
  Conf.}}, 2002, pp. 94--100.

\bibitem{Peeters2003}
G.~Peeters and X.~Rodet, ``{Signal-based Music Structure Discovery for Music
  Audio Summary Generation},'' in \emph{{Proc. of the 29th Intl. Computer Music
  Conf.}}, 2003, pp. 15--22.

\bibitem{Chu2000}
S.~Chu and B.~Logan, ``{Music Summary using Key Phrases},'' {Hewlett-Packard
  Cambridge Research Laboratory}, Tech. Rep., 2000.

\bibitem{Cooper2002}
M.~Cooper and J.~Foote, ``{Automatic Music Summarization via Similarity
  Analysis},'' in \emph{{Proc. of the 3rd ISMIR Conf.}}, 2002, pp. 81--85.

\bibitem{Glaczynski2011}
J.~Glaczynski and E.~Lukasik, ``{Automatic Music Summarization: A "Thumbnail"
  Approach},'' \emph{{Archives of Acoustics}}, vol.~36, no.~2, pp. 297--309,
  2011.

\bibitem{Bartsch2005}
M.~A. Bartsch and G.~H. Wakefield, ``{Audio Thumbnailing of Popular Music using
  Chroma-based Representations},'' \emph{{IEEE Trans. on Multimedia}}, vol.~7,
  no.~1, pp. 96--104, 2005.

\bibitem{Carbonell1998}
J.~Carbonell and J.~Goldstein, ``{The Use of MMR, Diversity-based Reranking for
  Reordering Documents and Producing Summaries},'' in \emph{{Proc. of the 21st
  Annual Intl. ACM SIGIR Conf. on Research and Development in Information
  Retrieval}}, 1998, pp. 335--336.

\bibitem{Erkan2004}
G.~Erkan and D.~R. Radev, ``{LexRank: Graph-based Lexical Centrality as
  Salience in Text Summarization},'' \emph{{Journal of Artificial Intelligence
  Research}}, vol.~22, pp. 457--479, 2004.

\bibitem{Landauer1997}
T.~K. Landauer and S.~T. Dutnais, ``{A solution to Plato’s problem: The
  latent semantic analysis theory of acquisition, induction, and representation
  of knowledge},'' \emph{{Psychological Review}}, vol. 104, no.~2, pp.
  211--240, 1997.

\bibitem{Zhu2007}
X.~Zhu, A.~B. Goldberg, J.~V. Gael, and D.~Andrzejewski, ``{Improving Diversity
  in Ranking using Absorbing Random Walks},'' in \emph{{Proc. of the 5th North
  American Chapter of the Association for Computational Linguistics - Human
  Language Technologies Conf.}}, 2007, pp. 97--104.

\bibitem{Ribeiro2011}
R.~Ribeiro and D.~M. de~Matos, ``{Revisiting Centrality-as-Relevance: Support
  Sets and Similarity as Geometric Proximity},'' \emph{{Journal of Artificial
  Intelligence Research}}, vol.~42, pp. 275--308, 2011.

\bibitem{Raposo2015}
F.~Raposo, R.~Ribeiro, and D.~M. de~Matos, ``{On the Application of Generic
  Summarization Algorithms to Music},'' \emph{{IEEE Signal Processing
  Letters}}, vol.~22, no.~1, pp. 26--30, 2015.

\bibitem{Xu2005}
C.~X. Xu, N.~C. Maddage, and X.~S. Shao, ``{Automatic Music Classification and
  Summarization},'' \emph{{IEEE Trans. on Speech and Audio Processing}},
  vol.~13, no.~3, pp. 441--450, 2005.

\bibitem{Vaizman2014}
Y.~Vaizman, B.~McFee, and G.~Lanckriet, ``{Codebook-based Audio Feature
  Representation for Music Information Retrieval},'' \emph{{IEEE/ACM Trans. on
  Audio, Speech and Language Processing}}, vol.~22, pp. 1483--1493, 2014.

\bibitem{Gong2001}
Y.~Gong and X.~Liu, ``{Generic Text Summarization Using Relevance Measure and
  Latent Semantic Analysis},'' in \emph{{Proc. of the 24th Annual Intl. ACM
  SIGIR Conf. on Research and Development in Information Retrieval}}, 2001, pp.
  19--25.

\bibitem{Steinberger2004}
J.~Steinberger and K.~Jezek, ``{Using Latent Semantic Analysis in Text
  Summarization and Summary Evaluation},'' in \emph{{Proc. of ISIM}}, 2004, pp.
  93--100.

\bibitem{Zechner2000}
K.~Zechner and A.~Waibel, ``{Minimizing Word Error Rate in Textual Summaries of
  Spoken Language},'' in \emph{{Proc. of the 1st North American Chapter of the
  Association for Computational Linguistics Conf.}}, 2000, pp. 186--193.

\bibitem{Murray2005}
G.~Murray, S.~Renals, and J.~Carletta, ``{Extractive Summarization of Meeting
  Recordings},'' in \emph{{Proc. of the 9th European Conf. on Speech
  Communication and Technology}}, 2005, pp. 593--596.

\bibitem{MIREX}
``{Music Information Retrieval Evaluation eXchange},''
  http://www.music-ir.org/mirex/wiki/MIREX\_HOME.

\bibitem{Wu2015}
M.-J. Wu and J.-S.~R. Jang, ``{Combining Acoustic and Multilevel Visual
  Features for Music Genre Classification},'' \emph{{ACM Trans. on Multimedia
  Computing, Communications and Applications}}, vol.~12, no.~1, pp.
  10:1--10:17, 2015.

\bibitem{Chang2011}
C.-C. Chang and C.-J. Lin, ``{LIBSVM: A Library for Support Vector Machines},''
  \emph{{ACM Trans. on Intelligent Systems and Technology}}, vol.~2, no.~3, pp.
  27:1--27:27, 2011.

\bibitem{deLeon2013}
F.~de~Leon and K.~Martinez, ``{Using Timbre Models for Audio Classification},''
  in \emph{{Submission to Audio Classification (Train/Test) Tasks of MIREX
  2013}}, 2013.

\bibitem{Curtin2013}
R.~R. Curtin, J.~R. Cline, N.~P. Slagle, W.~B. March, P.~Ram, N.~A. Mehta, and
  A.~G. Gray, ``{MLPACK: A Scalable C++ Machine Learning Library},''
  \emph{{Journal of Machine Learning Research}}, vol.~14, no.~1, pp. 801--805,
  2013.

\bibitem{Eyben2013}
F.~Eyben, F.~Weninger, F.~Gross, and B.~Schuller, ``{Recent Developments in
  openSMILE, the Munich Open-source Multimedia Feature Extractor},'' in
  \emph{{Proc. of the 21st ACM Intl. Conf. on Multimedia}}, 2013, pp. 835--838.

\bibitem{Sanderson2010}
C.~Sanderson, ``{Armadillo: An Open Source C++ Linear Algebra Library for Fast
  Prototyping and Computationally Intensive Experiments},'' {NICTA}, Tech.
  Rep., 2010.

\bibitem{Tzanetakis1999}
G.~Tzanetakis and P.~Cook, ``{MARSYAS: A Framework for Audio Analysis},''
  \emph{{Organised Sound}}, vol.~4, no.~3, pp. 169--175, 1999.

\bibitem{Weiss2010}
R.~Weiss and J.~P. Bello, ``{Identifying Repeated Patterns in Music Using
  Sparse Convolutive Non-Negative Matrix Factorization},'' in \emph{{Proc. of
  the 11th ISMIR Conf.}}, 2010, pp. 123--128.

\end{thebibliography}


\begin{IEEEbiography}[{\includegraphics[width=1in,height=1.25in,clip,keepaspectratio]{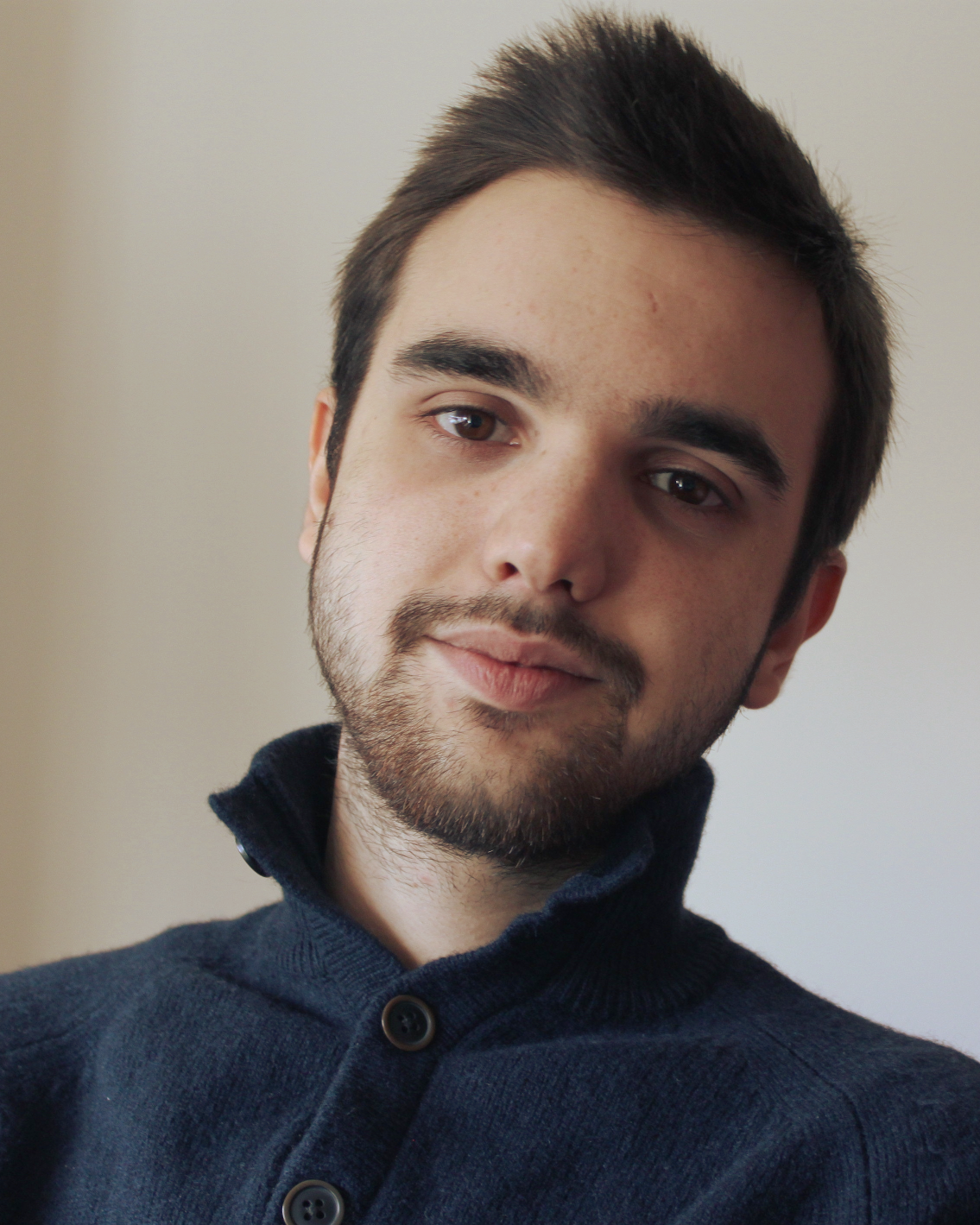}}]{Francisco
Raposo} graduated in Information Systems and Computer Engineering (2012) from
Instituto Superior Técnico (IST), Lisbon. He received a Masters Degree in
Information Systems and Computer Engineering (2014) (IST), on automatic music
summarization. He's currently pursuing a PhD course on Information Systems and
Computer Engineering. His research interests focus on music information
retrieval (MIR), music emotion recognition, and creative-MIR applications.
\end{IEEEbiography}

\begin{IEEEbiography}[{\includegraphics[width=1in,height=1.25in,clip,keepaspectratio]{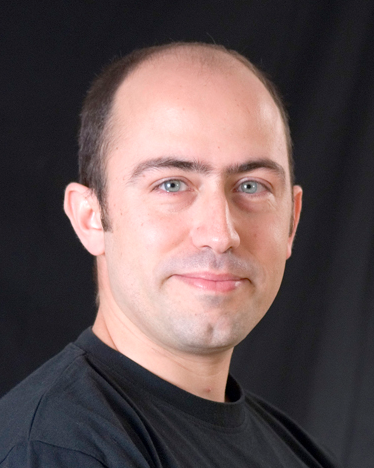}}]{Ricardo
Ribeiro} has a PhD (2011) in Information Systems and Computer Engineering and an
MSc (2003) in Electrical and Computer Engineering, both from Instituto Superior
Técnico, and a graduation degree (1996) in Mathematics/Computer Science from
Universidade da Beira Interior. His current research interests focus on
high-level information extraction from unrestricted text or speech, and
improving machine-learning techniques using domain-related information.
\end{IEEEbiography}

\begin{IEEEbiography}[{\includegraphics[width=1in,height=1.25in,clip,keepaspectratio]{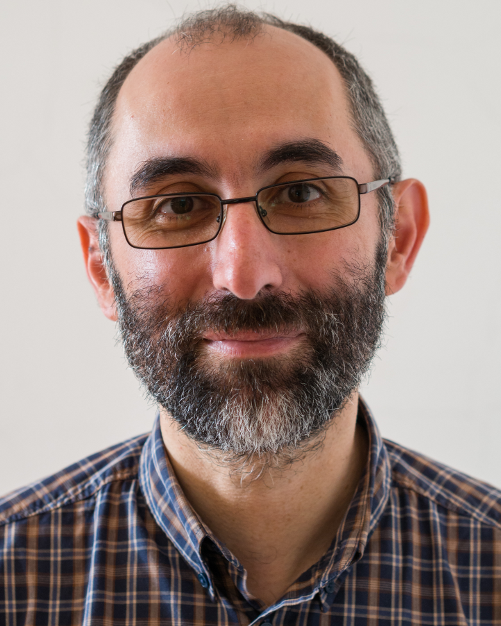}}]{David
Martins de Matos} graduated in Electrical and Computer Engineering (1990) from
Instituto Superior Técnico (IST), Lisbon. He received a Masters Degree in
Electrical and Computer Engineering (1995) (IST). He received a Doctor of
Engineering Degree in Systems and Computer Science (2005) (IST). His current
research interests focus on computational music processing, automatic
summarization and natural language generation, human-robot interaction, and
natural language semantics.
\end{IEEEbiography}

\end{document}